# Drop Formation from a Wettable Nozzle

Brian Chang, Gary Nave, and Sunghwan Jung*

*Department of Engineering Science and Mechanics,*
*Virginia Polytechnic Institute and State University, Blacksburg, VA 24061*

**Abstract:**

The process of drop formation from a nozzle can be seen in many natural systems and engineering applications. Here, we investigate the formation of a liquid droplet from a wettable nozzle. The behavior of a drop is complicated due to an interplay among gravity, capillary rise, viscous drag, and surface tension. In experiments, we observe that drops forming from a wettable nozzle initially climb the outer walls of the nozzle due to surface tension. Then, when the weight of the drops gradually increases, they finally fall due to gravity. By changing the parameters like the nozzle size and fluid flow rate, we have observed that different behaviors of the droplets. Such oscillatory behavior is characterized by a nonlinear equation that consists of capillary rise, viscous drag, and gravity. Two asymptotic solutions in the initial and later stages of drop formation are obtained and show good agreement with experimental observations.

*Corresponding author:
228 Norris Hall, Virginia Tech, Blacksburg, VA
telephone: 540-231-5146
fax: 540-231-4574
email: sunnyjsh@vt.edu

## I. Introduction

The mechanism of drops forming from a nozzle has been employed in many industrial applications [1-4]; i.e. ink-jet printing, spray cooling, emulsion formation, 3D micro-printing, and more. In most applications, uniform size distribution and fast production rate of droplets are required for producing high-quality results and reducing the operation time [5, 6]. The key parameters for designing such mechanisms are ejecting speed of fluid as well as physical shape and chemical property of the nozzle [1]. In particular, the motion of a slowly emitted liquid strongly depends on the nozzle-exit surface property i.e. a tea-pot effect [7, 8].

In the process of drop formation, there are two primary modes in which a drop can be produced. The first mode is in which the drop separates itself from the tip of a long fluid column when in jetting mode [9]. The other mode requires lower flow speed so that drops can detach themselves from the orifice, often called dripping mode [10]. Drop formation in the jetting mode has been extensively studied from the perspective of capillary stability of jet breakup[1]. The transition from dripping to jetting mode has been observed too[11, 12].

In the dripping mode, the size and shape of the drops become highly dependent on the nozzle-exit condition. However, previous studies have been made using a non-wettable orifice which provides a simple nozzle-exit condition on the contact line between the interface and the nozzle [10, 13]. The drop dynamics produced by a wettable surface has not been studied extensively. This study requires analysis of the interaction between the liquid and the outer walls of the nozzle in which capillary rise comes into play.

The natural phenomena of capillary rise has been studied for almost a century now [14, 15]. It is a common mechanism used by plants to transport water from the roots, through the stem, and into the leaves. Such capillary forces are also utilized by industrial systems; i.e., dying colors, driving ink in pens, and chromatography. The Washburn equation [15] is widely used as the governing equation that characterizes capillary rise in small tubes. We developed a similar equation that can be employed for capillary rise outside a nozzle.

In this article, we investigated the dynamics of drop formation from a wettable nozzle. In contrast to drops forming from a non-wettable nozzle, drops initially rise due to capillary force when emerging from a wettable nozzle. The drops then fall as soon as the weight is sufficiently large enough. In section II, the experimental setup and procedure is described. The theoretical model and its comparison with the experimental results are discussed in section III. In section IV, we discuss the conclusions of our findings and the future direction of this research.

## II. Experimental Methods

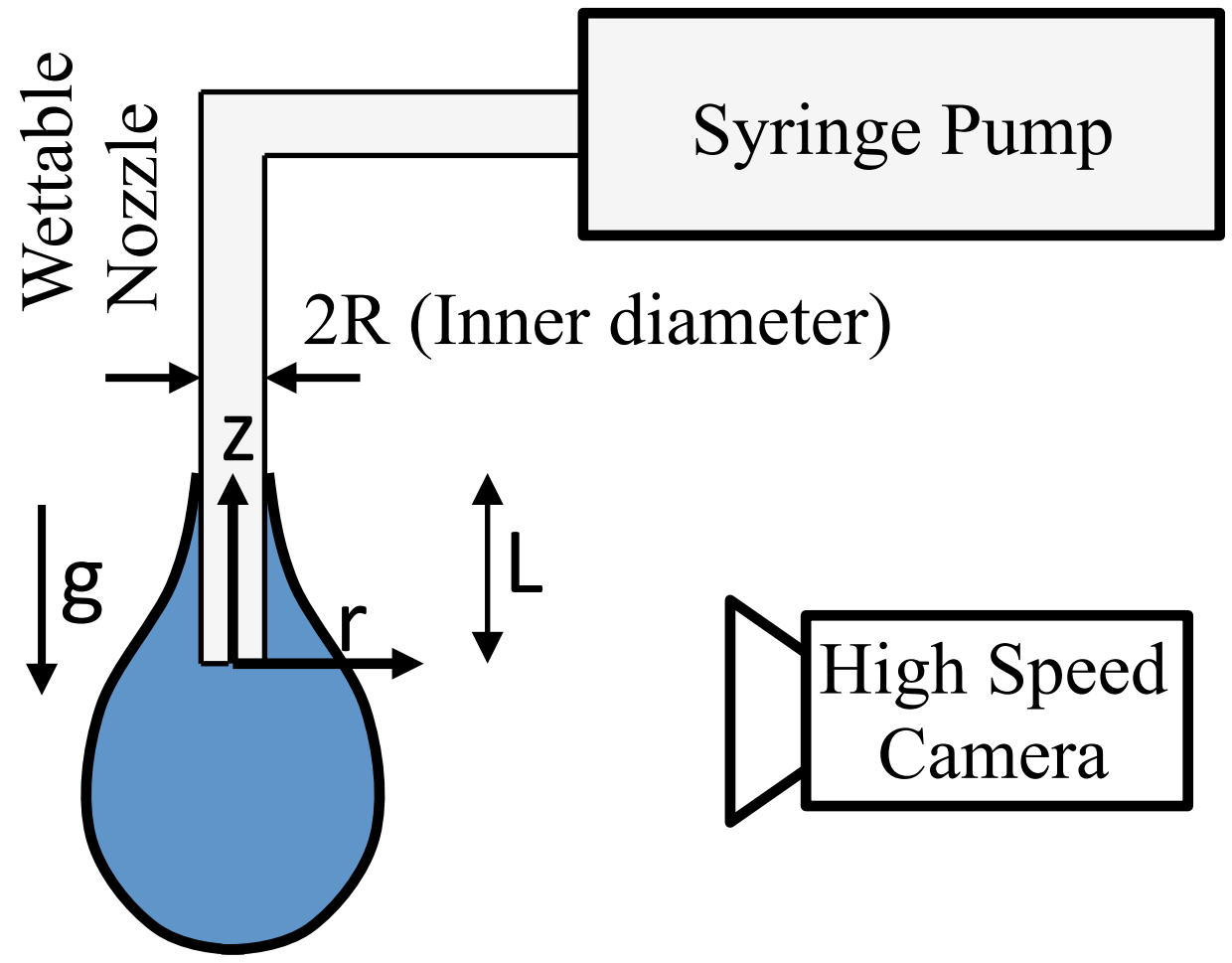

Figure 1. Schematics of the experimental setup.

The experimental setup was designed to visually analyze the behavior of droplets emerging from a wettable orifice. A schematic of the experimental apparatus is shown in Figure 1. By changing the flow rate, fluid viscosity, and tube diameter, we have tested the dynamics of drops forming from wettable nozzles. In order to vary the flow rates, the silicone fluid with a viscosity of 10 cSt was pumped by a syringe pump (New Era System; NE-1000) in the low velocities range (from 50 to 250 mL/hr with 50 mL/hr intervals) in order to avoid the jetting mode. The syringes used were 19, 20 and 21 gauges (corresponding inner diameters are R = 0.406, 0.324, and 0.292 mm).

The motion of the droplets was recorded with a high speed camera (Sony XDR_HR100). With high-contrasted movies, we were able to analyze the motion of the droplets using digital imaging processing in MATLAB. Then, the motion and shape of each drop was analyzed frame by frame and the centroidal position was calculated.

Details on the image analysis using the MATLAB code are as follows. Each individual frame was converted into a black and white image in order to identify the boundaries of the drop (not including the nozzle). After the boundaries were found, the code averaged the radius at each point along the vertical axis. This assumes an axi-symmetric shape about the vertical axis, but does not assume a spherical shape. This information allowed the program to calculate the centroid using the trapezoidal integration method.

Our experiments with low flow rates through a wettable orifice have been carried out. In the initial stages of drop formation, as shown in Figure 2, a small amount of fluid climbs the sides of the tube against gravity by wetting the outer surface of the nozzle. Gradually, the drop increases in weight and ultimately falls due to gravity.

Figure 3 shows the behavior of the centroid as a function of time with a fixed nozzle size. As the flow rate increases, the effect of capillary rise is less prevalent. The weight of the drop dominates the surface tension and viscous drag at high flow rates; the drop has less time to climb up the outer surface of the nozzle. Eventually, as we continue to increase the flow rate, the fluid does not make contact with the outer surface of the nozzle and goes into jetting mode. The centroid motion of drops is observed with different nozzle sizes at a fixed flux as shown in Figure 4. Initially, the drop climbs up in a similar fashion regardless of the nozzle size, but the later dropping behaviors are different. Overall, the period and maximum drop height decrease with decreasing the nozzle size.

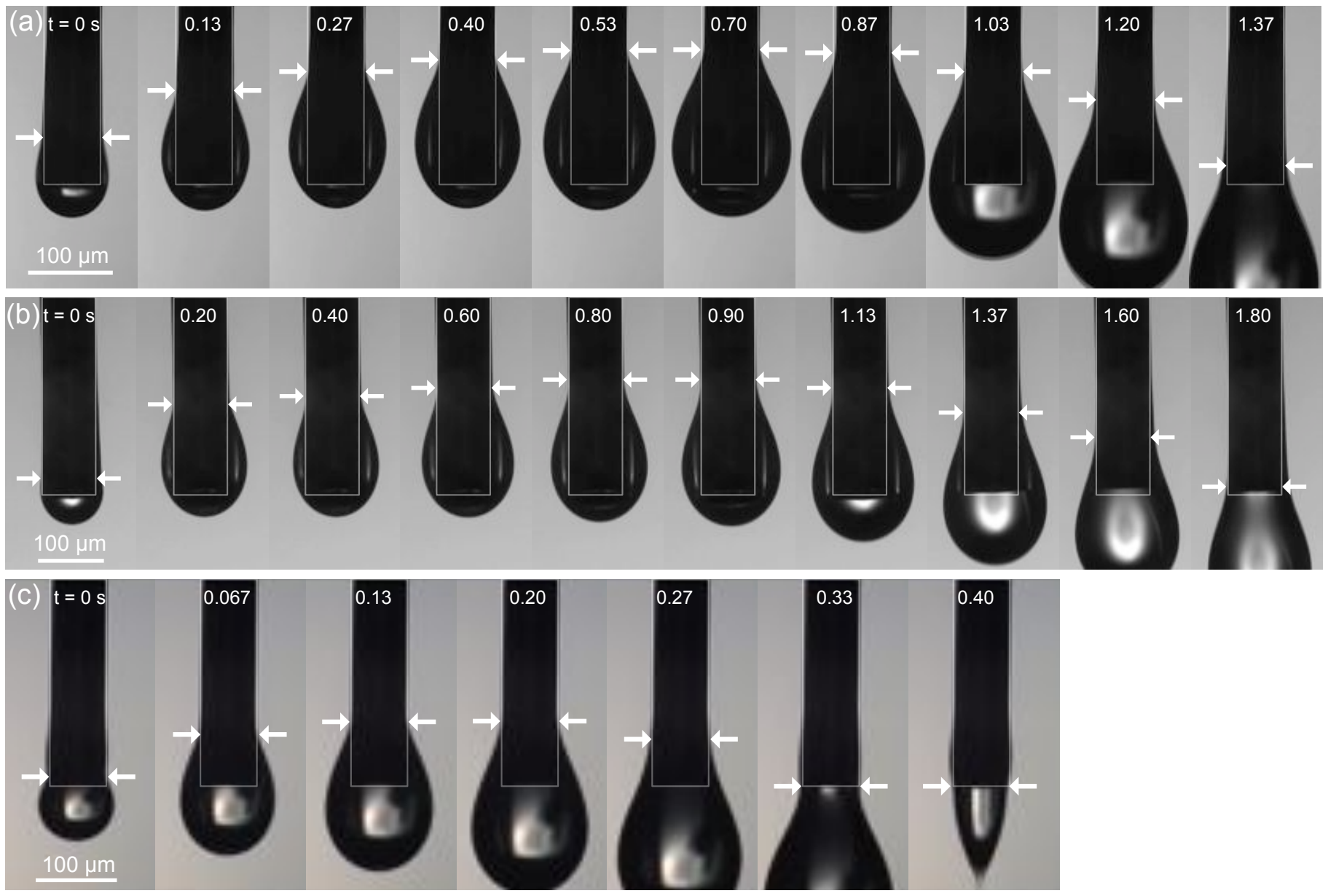

**Figure 2**. Drop formed from different nozzles with (a) R = 0.292 mm at 50 mL/hr, (b) R = 0.406 mm at 50 mL/hr, (c) R = 0.292 mm at 200 mL/hr.

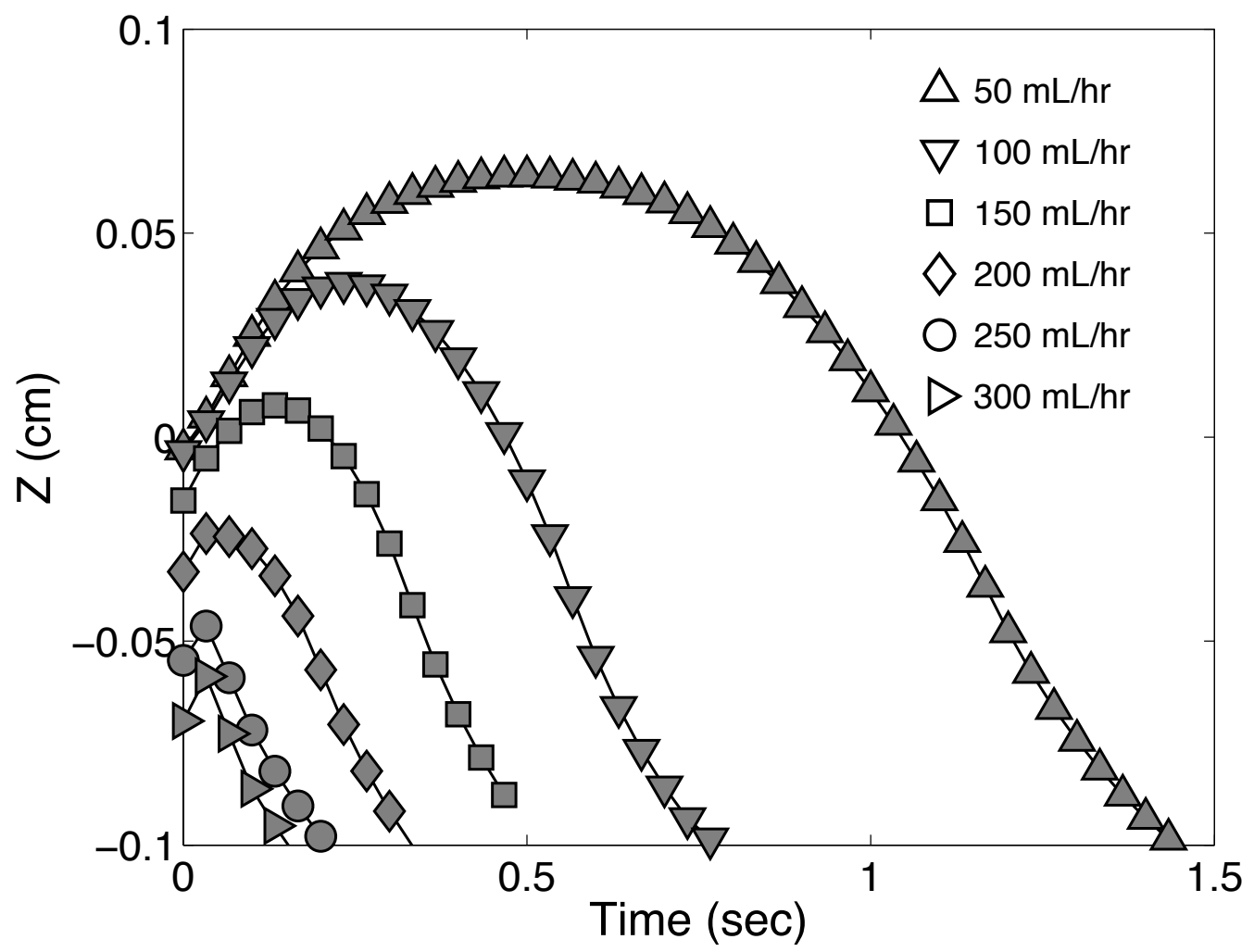

**Figure 3.** Centroidal motion of drop with different flow rates with R = 0.292 mm.

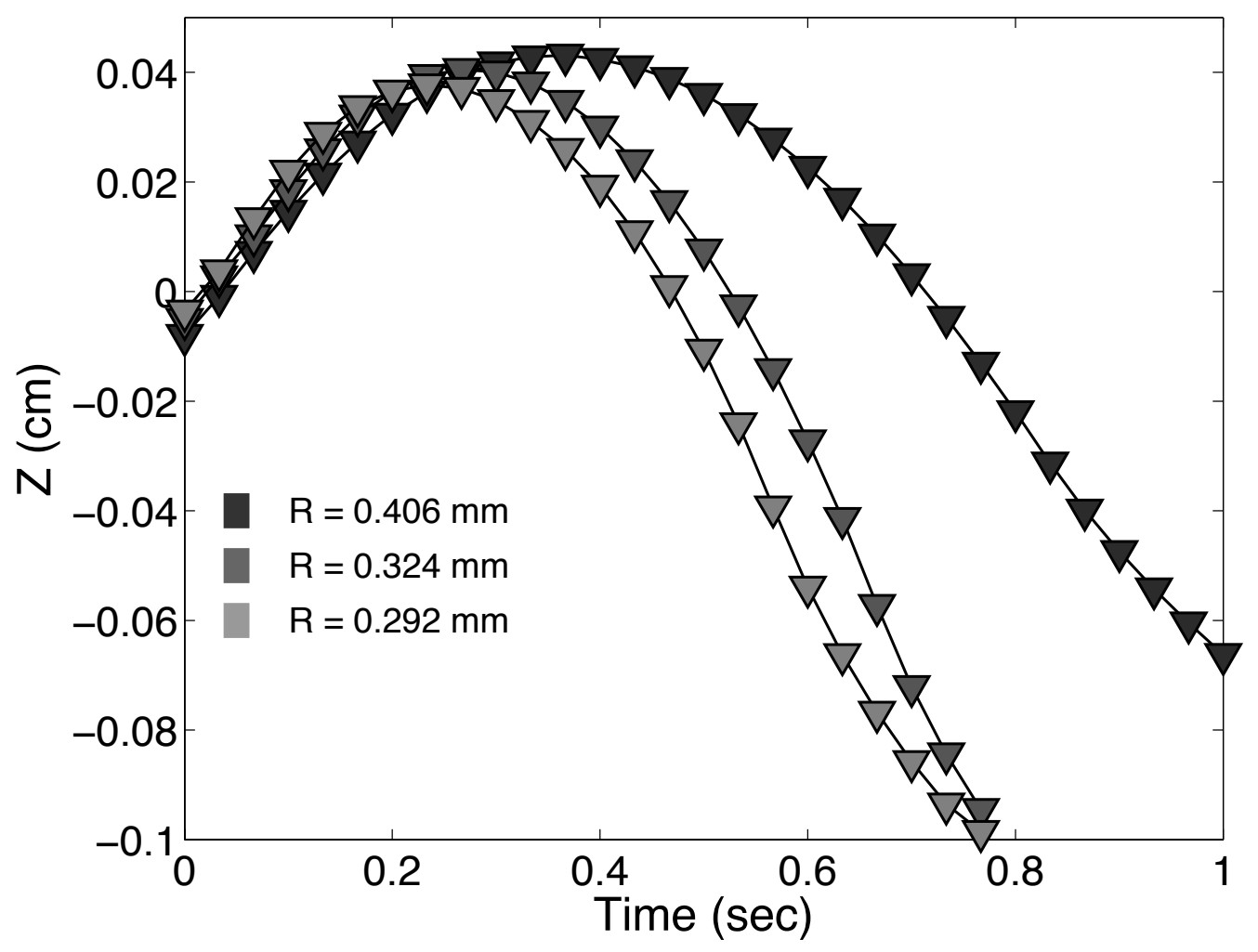

**Figure 4.** Centroidal motion of drop with different nozzle sizes with 100 mL/hr flow rate.

## III. Theoretical model

Surface tension, viscous drag and gravity are the main forces controlling the motion of the drop. For simplicity, the viscous dissipation associated with the displacement of the contact line is neglected. For the above system of a drop emerging from a nozzle, we express a governing equation as:

$$\frac{d(MV)}{dT} = \gamma(2\pi R)\cos(\theta_c) \pm 4\pi\mu VL - Mg$$

where $M$ is the fluid mass, $V$ the vertical velocity, $\theta_c$ the contact angle, and $L$ the length of the wetted nozzle. The fluid mass in a drop linearly increases with time by pumping a fluid at a constant rate at $M = \dot{M}(T\text{-}T_0)$, where $\dot{M}$ is the mass flux rate and $T_0$ is the initial time at the beginning of drop formation. By introducing non-dimensionalized parameters as $m = M/(\rho R^3)$, $t = T/(\rho R^3/\dot{M})$, $v = V/(\dot{M}/\rho R^2)$, and $l=L/R$, the governing equation becomes:

$$\frac{d(mv)}{dt} = \frac{2\pi\cos(\theta_c)}{We} \pm \frac{4\pi}{Re}vl - \frac{1}{Fr^2}m$$

Here, the Weber number is the ratio of inertial forces to the surface tension, defined as $We = \rho V^2 R / \gamma$, where $R$ is the characteristic size, $V$ the characteristic velocity ($V= \dot{M}/\rho R^2$), $\rho$ the fluid density, and $\gamma$ the surface tension. The capillary number ($Ca = \eta V/\gamma$, where $\eta$ is the fluid viscosity) is the ratio of viscous stress to surface tension. The Reynolds number is the ratio of inertia to viscous forces and is defined as $Re = \rho VR/\eta$. The Froude number is a ratio of inertial to gravitational forces and is defined as $Fr = V/(gR)^{1/2}$, $g$ being the acceleration due to gravity.

The beginning of the process takes place at a low Weber number and the surface tension plays a dominant role. It was observed that as the emerging fluid wets the outer wall of the nozzle, both the contact line and the drop are accelerated upwards by the capillary force. The drop is approximated as an annular shape with volume approximately proportional to $zR^2$, where $z$ is the vertical distance traveled by the drop. Then, the above equation in the low Weber limit becomes:

$$\rho\frac{d\left(z\frac{dz}{dt}\right)}{dt} = \frac{2\pi\cos(\theta_c)}{We} = const.$$

Assuming that $z \sim t^\alpha$, we find that $\alpha = 1$ at the beginning of the stage and the vertical position of drop increases as $z(t) \sim We^{-1/2}t$, which has the different scaling compared to Washburn's result describing capillary rise inside a tube ($z(t) = We^{-1/2}t^{1/2}$). The comparison between the Weber number and upward velocity ($dz/dt$) at the initial stage is shown in Figure 5. This comparison was typically made using the average velocity between the first three data points of each case. The dotted line in the figure is the rearrangement of $z(t) \sim We^{-1/2}t$ as $\log(We) \sim -2\log(dz/dt)$. The experimental data matches well with this predicted behavior. As the Weber number increases, equivalently as diameter increases or flow rate decreases, the upward velocity of the drop decreases along the predicted slope.

Later, the Froude number becomes large and the surface tension and viscous drag terms become negligible as the weight of the drop dominates the motion. The drop then slides downward due to gravity. After the moment of the highest vertical position ($t=t_m$), the fluid mass still increases as $m(t) = m(t_m)+(t-t_m)$ but its growth rate is smaller compared to the fluid mass $m(t_m)$ at the low pumping rate, and $z(t)$ starts from $z(t_m)$. The above equation by assuming the constant mass $m$ predicts the asymptotic solution of the vertical position as:

$$z(t) = z_m - Fr^{-2}t^2$$

By taking $dz/d(t^2)$, this equation becomes $dz/d(t^2) \sim Fr^{-2}$. This relation corresponds to the dotted line in Figure 6. The experiment showed that this prediction was accurate for the downward motion of the drop. By taking the absolute value of the average velocity of eight points in the later stage ($t = 2t_m$), the experimental data points are obtained and follow the predicted slope of -1/2 nicely, supporting our prediction. As the Froude number increases (as diameter decreases or flow rate increases), the downward velocity of the drop decreases along the predicted slope, as shown in Figure 6.

These asymptotic analyses predict the capillary-driven upward motion and the gravity-driven downward motion. The detailed dynamics can be obtained by solving the governing equation numerically.

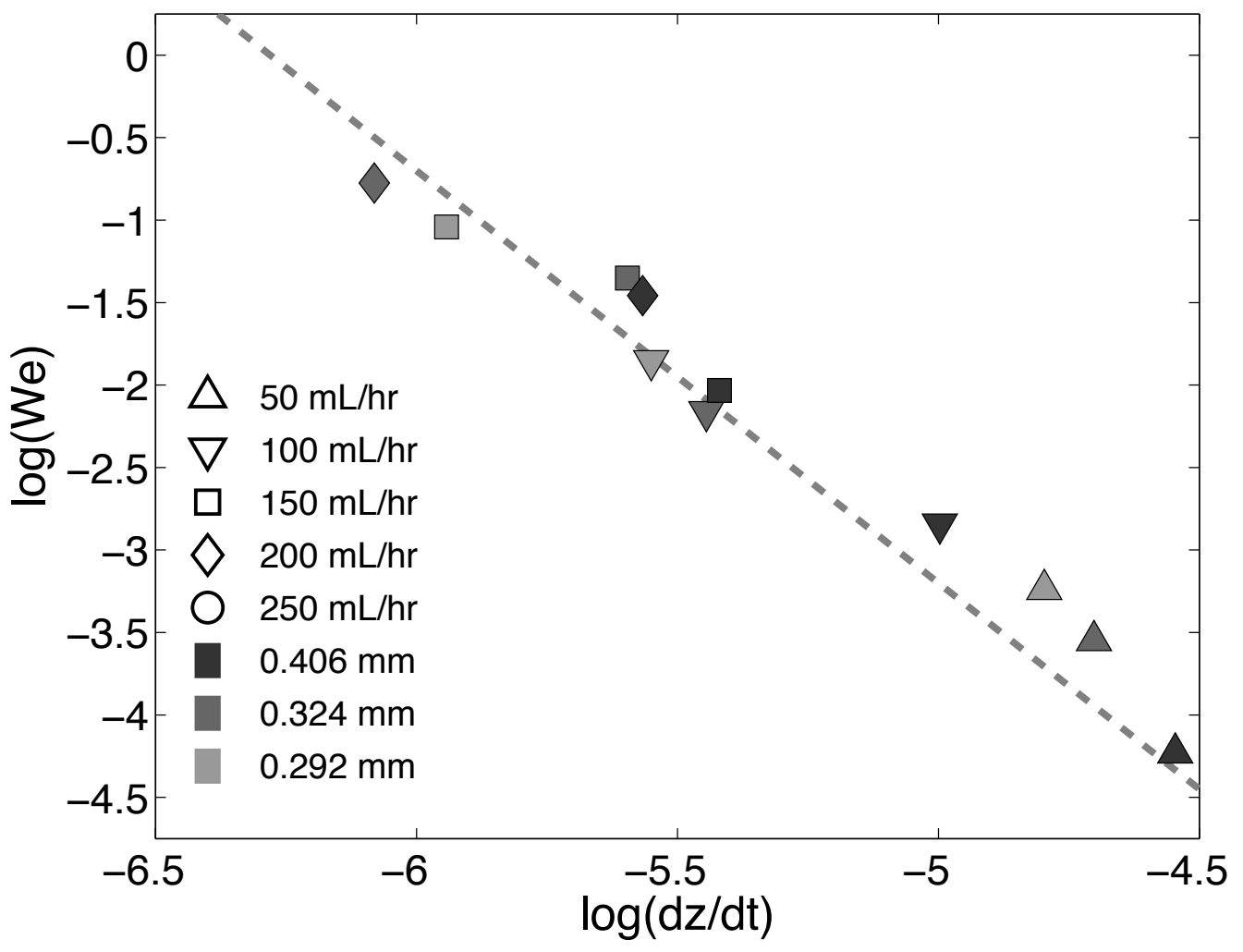

**Figure 5:** Weber number vs. drop velocity (*dz/dt*) with different radii and flow rates. The dotted line is from the asymptotic behavior $dz/dt \sim We^{-1/2}$ predicted from our calculation. The plotted points are the averaged velocity from the early stage of the droplet motion.

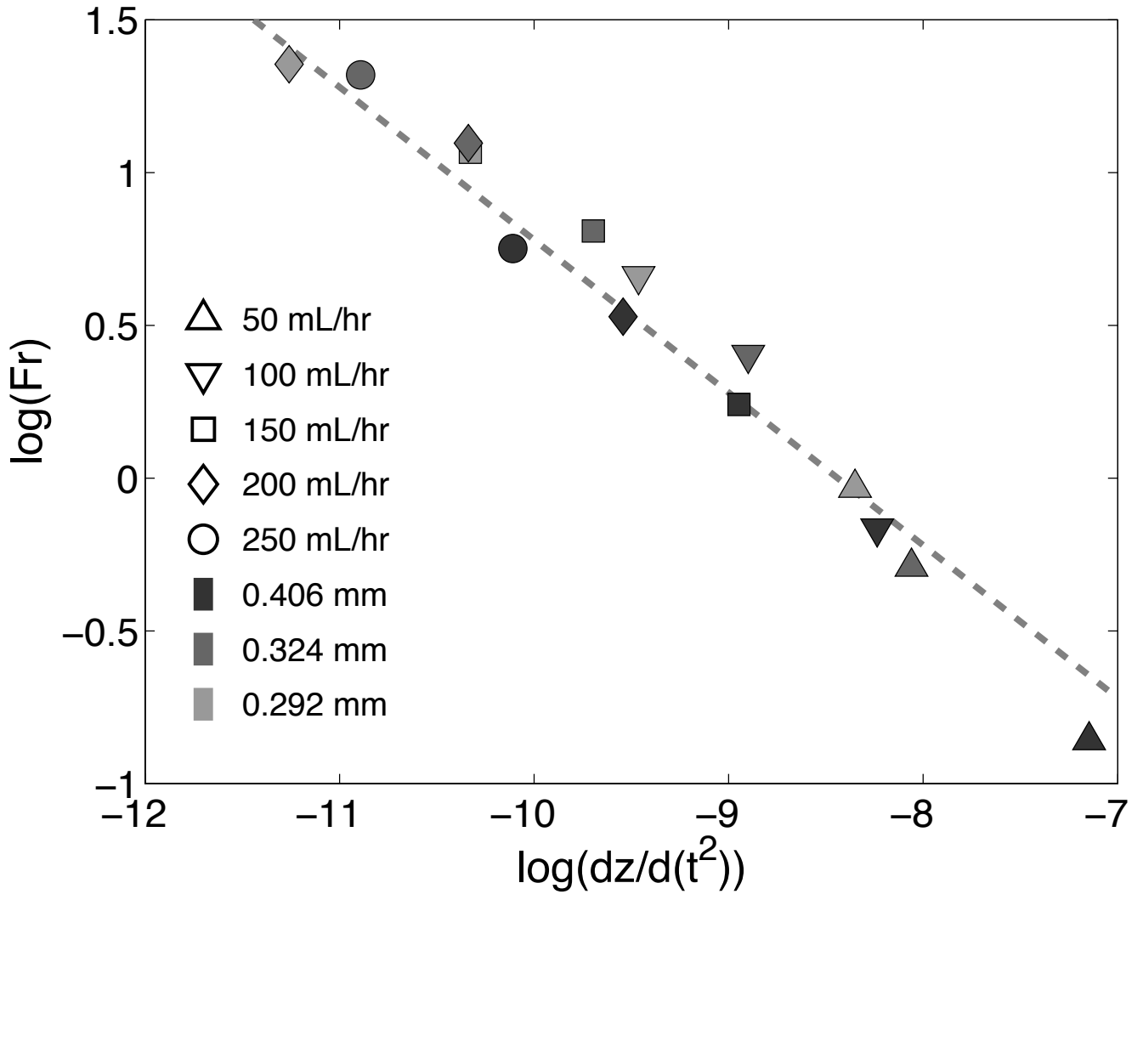

**Figure 6:** Froude number vs. drop motion (*dz/dt*$^2$) with different radii and flow rates. The dotted line is from the asymptotic behavior d*z*/d*t* $\sim Fr^{-2}$ predicted from our calculation. The points on the graph are taken from the average velocity near t = $2t_m$.

## IV. Conclusion

In this article, we investigated the process of drop formation from a wettable nozzle. In the process of forming a single droplet from a wettable orifice, the drop initially undergoes upward movement due to surface tension and then, once the droplet is sufficiently large, falls downward due to gravity. The third and final force acting on the droplet is viscous drag along the boundary between the droplet and the surface. Asymptotically, the drop starts out with an upward rise based on $z \sim t/\,We^{\frac{1}{2}}$, as it is more accurately approximated by a cylinder than a sphere, and finishes with a free fall and follows the behavior as $z \sim -\,t^2/Fr^2$. Experimental observations were made by varying diameter and volumetric flow rate in experiments, and are in good agreement with the asymptotic solutions found in the theoretical model. The determination of such a model for this behavior is also relevant in applications with flows under the electric field, such as microfluidics and ink-jet printer technology. In this context, we plan to combine the proposed experiments with electrowetting techniques in the future, in order to vary the surface tension while the drop is at the orifice.

## V. Acknowledgement

The author, S. Jung, celebrates the 60th birthday of Prof. Philip Morrison, a great advisor for guiding S. Jung into the wild world of science with his thoughtful (on-going) insights into seeking a theory of everything.

## VI. References

1. Bogy, D., *Drop formation in a circular liquid jet.* Annual Review of Fluid Mechanics, 1979. **11**(1): p. 207-228.
2. Hon, K., L. Li, and I. Hutchings, *Direct writing technology--Advances and developments.* CIRP Annals-Manufacturing Technology, 2008. **57**(2): p. 601-620.
3. Beale, J. and R. Reitz, *Modeling spray atomization with the Kelvin-Helmholtz/Rayleigh-Taylor hybrid model.* Atomization and sprays, 1999. **9**(6): p. 623-650.
4. Thorsen, T., et al., *Dynamic pattern formation in a vesicle-generating microfluidic device.* Physical review letters, 2001. **86**(18): p. 4163-4166.
5. GaÒan-Calvo, A.M., *Generation of steady liquid microthreads and micron-sized monodisperse sprays in gas streams.* Physical review letters, 1998. **80**(2): p. 285-288.
6. Basaran, O.A., *Small scale free surface flows with breakup: Drop formation and emerging applications.* AIChE Journal, 2002. **48**(9): p. 1842-1848.
7. Duez, C., et al., *Beating the teapot effect.* Arxiv preprint arXiv:0910.3306, 2009.
8. Reiner, M., *The teapot effect: a problem.* Physics Today, 1956. **9**: p. 16.
9. Rayleigh, L., *On the instability of jets.* Proceedings of the London Mathematical Society, 1878. **1**(1): p. 4.
10. Tate, T., *On the magnitude of a drop of liquid formed under different circumstances.* Philosophical Magazine Series 4, 1864. **27**(181): p. 176-180.

11. Clanet, C. and J. Lasheras, *Transition from dripping to jetting.* Journal of Fluid Mechanics, 1999. **383**: p. 307-326.
12. Utada, A., et al., *Absolute instability of a liquid jet in a coflowing stream.* Physical review letters, 2008. **100**(1): p. 14502.
13. Wilson, S., *The slow dripping of a viscous fluid.* Journal of Fluid Mechanics, 1988. **190**: p. 561-570.
14. Lucas, R., *Rate of capillary ascension of liquids.* Kolloid Z, 1918. **23**: p. 15-22.
15. Washburn, E., *The dynamics of capillary flow.* Physical Review, 1921. **17**(3): p. 273-283.